\documentclass[conference]{IEEEtran}
\IEEEoverridecommandlockouts
\usepackage{cite}
\usepackage{amsmath,amssymb,amsfonts}
\usepackage{algorithmic}
\usepackage{graphicx}
\usepackage{textcomp}
\usepackage{xcolor}
\usepackage{booktabs}
\usepackage{multirow}
\def\BibTeX{{\rm B\kern-.05em{\sc i\kern-.025em b}\kern-.08em
    T\kern-.1667em\lower.7ex\hbox{E}\kern-.125emX}}
\def\mycopyrightnotice{
{\hfill \footnotesize 978-1-6654-0126-5/21/\$31.00 \copyright 2021 IEEE\hfill}
}
\makeatother
\begin{document}

\title{CoCo DistillNet: a Cross-layer Correlation Distillation Network for Pathological Gastric Cancer Segmentation

\thanks{$\ast$ These authors contributed equally to this work.}
\thanks{$\dagger$ Corresponding authors.}
}
\author{Wenxuan Zou\textsuperscript{1,$\ast$}, Xingqun Qi\textsuperscript{1,$\ast$},  Zhuojie Wu\textsuperscript{1}, Zijian Wang\textsuperscript{1},  Muyi Sun\textsuperscript{2,$\dagger$}, Caifeng Shan\textsuperscript{3,4}\\
\textsuperscript{1}School of Automation, Beijing University of Posts and Telecommunications, Beijing, China\\
\textsuperscript{2}Center for Research on Intelligent Perception and Computing, NLPR, CASIA, Beijing, China\\
\textsuperscript{3}Shandong University of Science and Technology, Qingdao 266590, China\\
\textsuperscript{4}Artificial Intelligence Research, CAS (CAS-AIR), Beijing 100190, China\\

{\tt\small \{zouwenxuan,zhuojiewu,wangzijianbupt,xingqunqi\}@bupt.edu.cn} \\
{\tt\small  muyi.sun@cripac.ia.ac.cn, caifeng.shan@gmail.com } \\
}
\maketitle

\begin{abstract}
In recent years, deep convolutional neural networks have made significant advances in pathology image segmentation. However, pathology image segmentation encounters with a dilemma in which the higher-performance networks generally require more computational resources and storage. This phenomenon limits the employment of high-accuracy networks in real scenes due to the inherent high-resolution of pathological images. To tackle this problem, we propose CoCo DistillNet, a novel Cross-layer Correlation (CoCo) knowledge distillation network for pathological gastric cancer segmentation. Knowledge distillation, a general technique which aims at improving the performance of a compact network through knowledge transfer from a cumbersome network. Concretely, our CoCo DistillNet models the correlations of channel-mixed spatial similarity between different layers and then transfers this knowledge from a pre-trained cumbersome teacher network to a non-trained compact student network. In addition, we also utilize the adversarial learning strategy to further prompt the distilling procedure which is called Adversarial Distillation (AD). Furthermore, to stabilize our training procedure, we make the use of the unsupervised Paraphraser Module (PM) to boost the knowledge paraphrase in the teacher network. As a result, extensive experiments conducted on the Gastric Cancer Segmentation Dataset demonstrate the prominent ability of CoCo DistillNet which achieves state-of-the-art performance.
\end{abstract}

\let\thefootnote\relax\footnotetext{\mycopyrightnotice}
\section{Introduction}
In clinic medicine, pathology image segmentation serves as a beneficial measure to help pathologists diagnosis the region of cancerization, tumors and so on. Recently,  with the development of deep learning, various delicate Convolution Neural Networks (CNNs) are devised and applied in the pathology image segmentation. Simultaneously, such intricate networks exhibit state-of-the-art performance in the area of 
automatic pathology image segmentation \cite{ronneberger2015u,tokunaga2019adaptive}.

Nevertheless, high-performance networks usually involve complicated, deep, and wide network architecture with numerous parameters. Furthermore, the size of Whole-Side Images (WSI) (e.g.
100,000$\times$50,000 pixels) which are commonly used in digital pathology is over 100 times a natural image \cite{tokunaga2019adaptive}. However, such high-resolution images cannot be sent into CNNs as inputs directly. To this end, previous researchers take the patch-based strategy as regular approach\cite{tokunaga2019adaptive} to handle this issues. Nevertheless, this strategy aggravates the requirement of memory and augments inference time. As drawbacks aforementioned, training an
appropriate network with high-performance and low consumption, as well as loading the model into embedded devices is nearly impractical, taking account of limited computation resources and storage of the medical devices in real scenes.

There have been various endeavors which engage in conquering the computation limitation and memory constraint in portable or edge devices. Model compression is one of the main directions which could be roughy categorized into three classes: network
quantization\cite{hubara2016binarized,yang2019quantization}, network pruning\cite{lecun1990optimal,han2015learning}, and knowledge distillation\cite{hinton2015distilling,romero2014fitnets,tung2019similarity,liu2019structured}. The purpose of network quantization is to transform the weights and features into low bitwidth integer types. Network pruning can compress the size of networks by cutting off some non-essential weights or neurons. Knowledge distillation is conducted to transfer
the knowledge of a pre-trained cumbersome teacher network to a compact student network in which the knowledge carries more inter-class and intra-class information than labels. However, network quantization exhibits poor performance in the large CNNs\cite{wu2016quantized}, while the network pruning criteria demands hand-craft sensitivity for layers\cite{cheng2018model}. Furthermore, it is worth noting that network pruning or quantization need specific
devices or implementations for accelerating CNNs. On the contrary, knowledge distillation can be applied in various network
architectures (e.g. collaboration with lightweight CNNs). Meanwhile, this method appears high efficiency with high accuracy and low consumption, as the student network can acquire knowledge from the teacher network that significantly outperforms the student. In addition, knowledge distillation does not need extra parameters whether in the process of training or inference.

In this paper, we adopt knowledge distillation to circumvent the tradeoff between accuracy and efficiency in pathology segmentation. Initially, knowledge distillation is defined as a method that a student network mimics the ‘soft targets’ from a teacher network, i.e. the prediction of teacher network can supply extra supervision to train the student network\cite{hinton2015distilling}. Next, knowledge distillation is extended to approximate the middle features of CNNs\cite{romero2014fitnets}. However, different network architectures are prone to extract different features in the middle layer of networks. Consequently, directly mimicking the middle features between student and teacher is inappropriate when student network has different architecture from the teacher network\cite{tung2019similarity}. Subsequently, many researchers study the similarity of inter-class and intra-class in the image classification task\cite{tung2019similarity,peng2019correlation,chen2020improving}. Considering the structural property of image segmentation, i.e. the relations among pixels in a feature map, researchers tend to transfer pixel-wise similarity of the teacher network to guide the training of the student network\cite{liu2019structured,hou2020inter,he2019knowledge,wang2020intra}.

The state-of-the-art networks have a common property, which tend to extract low-level representations in shallow layers as well as high-level representations in depth layers. However, recent knowledge distillation methods for image segmentation pay less attention to the correlation between low-level and high-level representations. Motivated by the Flow of Solution Procedure (FSP matrix)\cite{yim2017gift}, we propose a novel distillation method that defines Cross-layer Correlations (CoCo) as transferred knowledge. However, direct computation of the FSP matrix from two layers is a strict constraint which is hard to mimic well between student and teacher networks. At the same time, considering the structural property that pathology images exist pixel-wise similarity, the variation of cross-layer spatial similarity is employed to measure the correlation from different layers. Specifically, the channel-wise information in the teacher network can facilitate the performance of the student network\cite{shu2020channel}. Therefore, we mix channel-wise information into the spatial similarity which can further improve the performance of pathological gastric cancer segmentation. Also, inspired by using conditional Generative Adversarial Network in knowledge distillation\cite{xu2017training}, we utilize the adversarial learning strategy to once more improve the efficiency of our method. Furthermore, in the training procedure, we discover the paraphraser module\cite{kim2018paraphrasing} in an unsupervised way which can stable the process of knowledge transfer between teacher and student network.

Our main contributions can be summarized as follows:

$\bullet$ We propose CoCo DistillNet, a novel Cross-layer Correlation Distillation Network for pathological gastric cancer segmentation.

$\bullet$ 
The cross-layer correlation of channel-mixed spatial similarity is designed as a structural and robust knowledge which are transferred from teacher network to student network.

$\bullet$ Adversarial Distillation is utilized to further improve the performance through a training style, and the unsupervised Paraphraser Module is exploited to stabilize our training procedure.

$\bullet$ 
Extensive experiments are conducted on the Gastric Cancer Segmentation Dataset. Both
qualitative and quantitative results validate the effectiveness of the proposed knowledge distillation method.

\section{Related Work}
\subsection{Pathology Image Segmentation}
Pathology image segmentation dedicate to prompt diagnostic efficiency and accuracy
by automatic labeling a corresponding category to each pixel in an image. Pathology image segmentation has witnessed impressive progress with the boost of deep neural networks. Ronneberger \emph{et al.}\cite{ronneberger2015u} presented U-net that consists of an encoder to extract context and a decoder to enable precise segmentation. For accurate diagnosis of interstitial lung diseases, Anthimopoulos \emph{et al.}\cite{anthimopoulos2018semantic} proposed a deep convolutional neural network which consists of dilated convolution. Bi\emph{et al.}\cite{bi2017stacked} designed a new stacked fully convolutional network with multi-channel learning and achieved great advances in colorectal cancer segmentation. Chan \emph{et al.}\cite{chan2019histosegnet} adopted a patch-based CNN in colon glands cancer segmentation, named HistoSegnet, which appears excellent performance and generalization. However, the inherent high-resolution pathological images prohibit deploying the networks aforementioned with millions of parameters in real scenes. Subsequently, it is urgent to propose a lightweight model.
\subsection{Lightweight model in medical image analysis}
Lightweight model aims to deploy networks which have high performance and low 
consumption in real scenes. In medical image analysis, there are various efforts to devise the lightweight model, which include designing lightweight networks, network pruning, network quantization, and knowledge distillation. A lightweight Spatial Attention U-net (SA U-net) was proposed by Guo \emph{et al.}\cite{guo2021sa} for retinal vessel segmentation. Fernandes \emph{et al.}\cite{fernandes2020automatic} presented an automatic generating compact DNN used for medical image diagnostic with the help of network pruning. Zhang \emph{et.al}\cite{zhang2021medq} exploited network quantization technology to design a novel compact framework which obtain high effectiveness in several public 3D medical segmentation datasets.
In the next, a few researchers utilized knowledge distillation to deal with the problem of efficiency in medical image analysis. Ho \emph{et al.}\cite{ho2020utilizing} proposed a self-training knowledge distillation method to improve the accuracy in chest X-ray images classification. Li \emph{et al.}\cite{li2020towards} utilized online mutual knowledge distillation to improve the performance of CT image segmentation. Inspired by previous researches, we firstly introduce knowledge distillation in pathology image segmentation.
\begin{figure*}[t]
\begin{center}
\includegraphics[width=\linewidth]{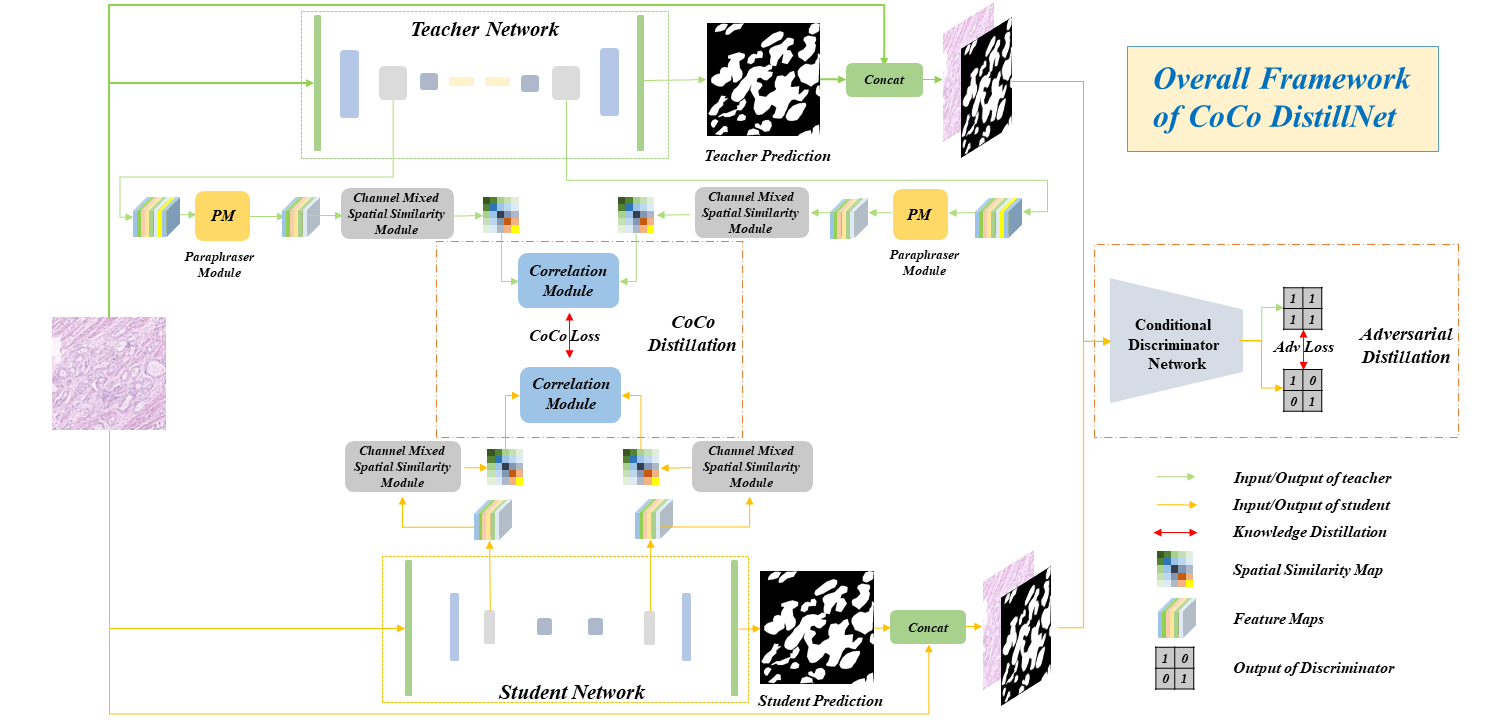}
\end{center}
   \caption{The overall framework of CoCo DistillNet. Channel Mixed Spatial Similarity Module is committed to extract the pixel-wise similarity. Meanwhile, it could also assimilate channel-wise information. Correlation Module is utilized to compute the cross-layer correlation.The student network is supervised by CoCo Distillation and Adversarial Distillation. Paraphraser Module aims to paraphrase the knowledge from teacher network.}
\label{fig:framework}
\end{figure*}
\subsection{Knowledge Distillation}The target of knowledge distillation is to obtain a student network that has competitive or even a superior performance compared with the teacher network. Initially, Hinton \emph{et al.}\cite{hinton2015distilling} regarded the 'soft target' of teacher network predictions as knowledge. Next, Ballas \emph{et al.}\cite{romero2014fitnets} proposed the hints from the intermediate feature map of the teacher network, which can also improve the performance of the student network. Subsequently, the hints from the intermediate feature map were further extended. Tung \emph{et al.}\cite{tung2019similarity} raised a 
novel knowledge distillation method which is to preserve pairwise similarity in a batch. Yim \emph{et al.}\cite{yim2017gift} treated the Flow of Solution Procedure 
(FSP matrix) as transferred knowledge and demonstrated the effectiveness in the CIFAR-100 dataset. He \emph{et al.}\cite{he2019knowledge} presented a method that minimizes the difference of pixel-wise affinity between teacher and student. Liu \emph{et al.}\cite{liu2019structured} investigate the structured knowledge in semantic segmentation, and then exam the effectiveness of two structured distillation schemes, which are pairwise distillation and holistic distillation. Motivated by preceding studies, we consider the cross-layer correlation for knowledge transfer.
\section{Method}
Pathology image segmentation aims to predict a corresponding class to each pixel in a given pathology image $X$. Given a pre-trained teacher network $T$ and a non-trained student network $S$, our CoCo DistillNet engages in improving the performance of $S$ supervised by the knowledge from $T$. The outputs of $S$ and $T$ are noted as $Y_{s}=S(X)$ and $Y_{t}=T(X)$. we set $F_{s}^{l_{s}^{j}}\in R^{ c\times h\times w} (j=1,\cdots,i,\cdots,n+i,\cdots,2n)$, where $c$ is the number of channels, $h \times w$ is the spatial size, and n is the number of the selected all $(l_{s}^{i},l_{s}^{n+i})$ pair layers, representing the feature maps of particular layer $l_{s}^{j}$ produced by $S$. $F_{t}^{l_{t}^{j}}\in R ^ {{c}'\times h\times w}(j=1,\cdots,i,\cdots,n+i,\cdots,2n)$ denote the feature maps of particular layer $l_{t}^{j}$ produced by $T$, where layer $l_{t}^{j}$ is determined  by the size of $F_{s}^{l_{s}^{j}}$. Note that ${{c}'}$ is not necessarily to equal ${c}$ due to the use of Parapharser Module. In this paper, we set n=1.
\subsection{Overview}
As illustrated in the Fig.~\ref{fig:framework}, the pathology image $X$ is input to the student network $S$ and the teacher network $T$, simultaneously. Given the feature maps $F_{t}^{l_{t}^{i}}$ from layer $l_{t}^{i}$, the PM produces new feature maps ${F_{t}^{l_{t}^{i}}}'$  which can be easily comprehended by student network $S$. Furthermore, the Channel Mixed Spatial Similarity Module is exploited to acquire the channel-mixed spatial similarity hidden in ${F_{t}^{l_{t}^{i}}}'$ and $F_{s}^{l_{s}^{i}}$. Subsequently, the Correlation Module measures the correlation of the channel-mixed spatial similarity between ${F_{t}^{l_{t}^{i}}}'$ and ${F_{t}^{l_{t}^{n+i}}}'$; $F_{s}^{l_{s}^{i}}$ and $F_{s}^{l_{s}^{n+i}}$, respectively. For transferring the cross-layer correlation knowledge from $T$ to $S$, we introduce the CoCo Loss to reduce the discrepancy of cross-layer correlation between $S$ and $T$. Furthermore, the Adversarial Distillation is employed to enforce $Y_{s}$ to mimic $Y_{t}$. At last, the student network $S$ trained by our CoCo DistillNet can achieve competitive performance without any extra parameters and memory.  
\subsection{Cross-layer Correlation Distillation}
\begin{figure}[t]
\begin{center}
\includegraphics[width=\linewidth]{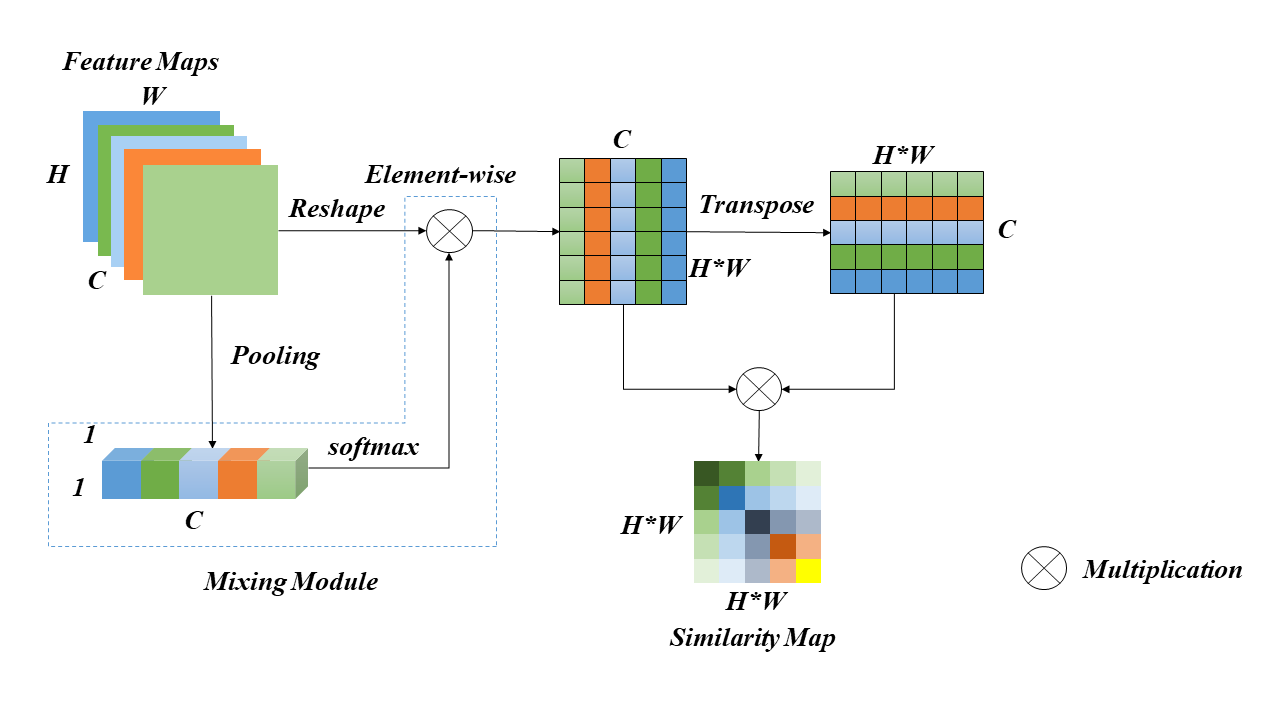}
\end{center}
   \caption{The illustration of Channel Mixed Spatial Similarity Module. Mixing Module is proposed to assimilate channel-wise information into spatial similarity maps.}
\label{fig:Channel mixed Spatial Similarity Module}
\end{figure}

\emph{1) Channel Mixed Spatial Similarity Module:} Considering the pixel-wise similarity in pathology images, it is crucial to introduce the spatial similarity in pathology images segmentation. Furthermore, we propose the channel-mixed spatial similarity as each channel tends to encode different categories information in the feature maps. Concretely, as depicted in the Fig.~\ref{fig:Channel mixed Spatial Similarity Module}, we use global max pooling to generate channel-wise information $A=[a_{1},a_{2},\cdots ,a_{c}]\in R^{c\times 1\times 1}$. Meanwhile, we reshape $F_{(\cdot)}^{l_{(\cdot)}^{i}}$ to $G_{(\cdot)}^{l_{(\cdot)}^{i}}\in R^{hw\times c}$ and exploit a softmax layer to normalize $A$. Next, we mix the channel-wise information ${A}$ into $G_{(\cdot)}^{l_{(\cdot)}^{i}}$ to calculate ${G_{(\cdot)}^{l_{(\cdot)}^{i}}}'$ , 
which refers to a $hw\times c$ matrix. The computation could be formulated as:
\begin{align}
{g_{ij}}'
=
g_{ij}
\frac{\exp (a_{j})}{\sum_{k=1}^{c} \exp(a_{k})}
\end{align}
where ${g_{ij}}'\in {G_{(\cdot)}^{l_{(\cdot)}^{i}}}'$. Subsequently, we defined the channel-mixed spatial similarity map as $M_{(\cdot)}^{l_{(\cdot)}^{i}}\in R^{hw\times hw}$: 
\begin{align}
m_{ij}= \sum_{k=1}^{c}
\frac{{g}'_{ik}{g}''_{kj}}{\sqrt{\sum_{z=1}^{c} {g}'_{iz} }\sqrt{\sum_{z=1}^{c} {g}''_{zj} }}
\end{align}
where $m_{ij}\in M_{(\cdot)}^{l_{(\cdot)}^{i}}$ refers to the similarity between i-th pixel and j-th pixel in the feature maps $F_{(\cdot)}^{l_{(\cdot)}^{i}}$, and ${g}''_{kj}$ is the element of $({G_{(\cdot)}^{l_{(\cdot)}^{i}}}')^{T}$.    

\begin{figure}[ht]
\begin{center}
\includegraphics[width=\linewidth]{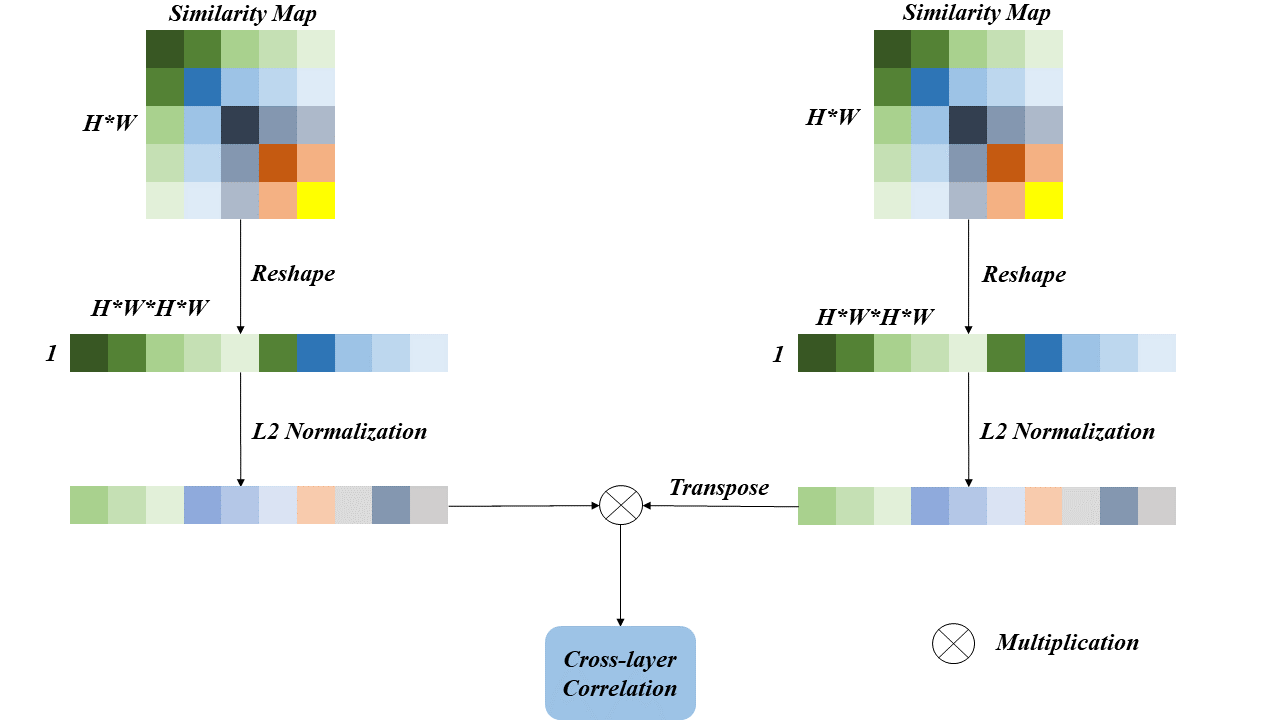}
\end{center}
   \caption{The pipeline of Correlation Module. The cross-layer correlation is measured by similarity maps from different layers.}
\label{fig:Correlation Module}
\end{figure}

\emph{2) Correlation Module:} 
We propose the Correlation Module to measure the correlation between low-level representations and high-level representations. As shown in Fig.~\ref{fig:Correlation Module}, we utilize $M_{(\cdot)}^{l_{(\cdot)}^{i}}$ and $M_{(\cdot)}^{l_{(\cdot)}^{n+i}}$ to calculate the correlation $\phi(\cdot,\cdot)$ of representations between $l_{(\cdot)}^{i}$ and $l_{(\cdot)}^{n+i}$. The $\phi(\cdot,\cdot)$ can be formulated as:
\begin{align}
\phi 
(F_{(\cdot)}^{l_{(\cdot)}^{i}},F_{(\cdot)}^{l_{(\cdot)}^{n+i}})
=
\frac{Q_{(\cdot)}^{l_{(\cdot)}^{i}}(Q_{(\cdot)}^{l_{(\cdot)}^{n+i}})^{T}}
{\left \| Q_{(\cdot)}^{l_{(\cdot)}^{i}} \right \|_{2}\cdot\left \| Q_{(\cdot)}^{l_{(\cdot)}^{n+i}} \right \|_{2}}
\end{align}
where $Q_{(\cdot)}^{l_{(\cdot)}^{i}}\in R^{1\times hwhw}$ is the reshaping of the $M_{(\cdot)}^{l_{(\cdot)}^{i}}$.

Next, we use Euclidean distance to minimize the discrepancy of the cross-layer correlation between teacher network and student network. 
We named this constraint as CoCo knowledge distillation loss function $\mathcal{L}_{coco}$ which could be presented as :
\begin{align}
    \mathcal{L} _{coco}=\frac{1}{n}\sum_{i=1}^{n}
    \left \| \phi 
(F_{t}^{l_{t}^{i}},F_{t}^{l_{t}^{n+i}}) -\phi 
(F_{s}^{l_{s}^{i}},F_{s}^{l_{s}^{n+i}})\right \|_{2}^{2}
\end{align}

\begin{figure}[ht]
\begin{center}
\includegraphics[width=\linewidth]{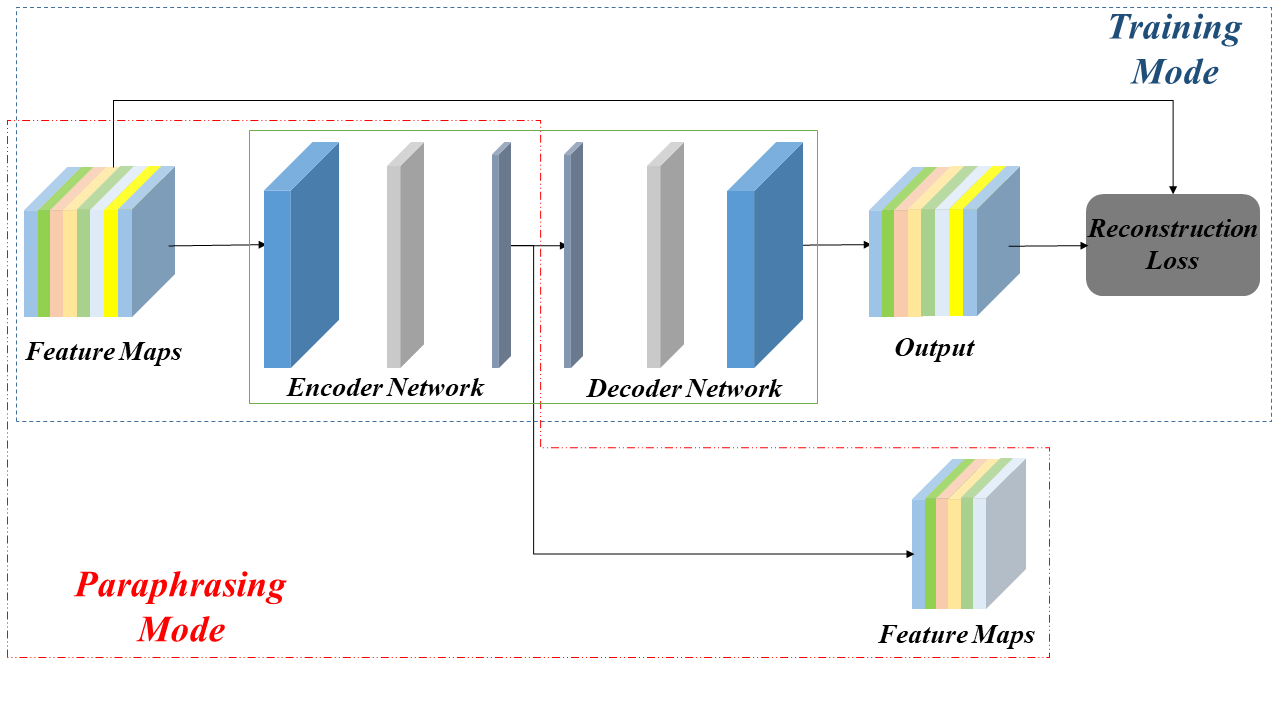}
\end{center}
   \caption{The overall architecture of Paraphraser Module. 
   The Training Mode is utilized to train the encoder and decoder network. In the Paraphrasing Mode, we only employ the encode network to paraphrase knowledge from teacher.}
\label{fig:PM}
\end{figure}

\emph{3) Paraphraser Module:}
In the training procedure,  the $\mathcal{L}_{coco}$ is difficult to converge stably and sufficiently. It indicates that the knowledge from $T$ is hardly transferred to $S$. Empirically, the knowledge embedded in $T$ can not be comprehended well by $S$ due to the feature distribution gap caused by mapping of $T$ and $S$. Motivated by \cite{kim2018paraphrasing}, we exploit the Paraphraser Module to embed the feature maps $F_{t}^{l_{t}^{i}}$ into the same feature space as $F_{s}^{l_{s}^{i}}$ without the loss of necessary knowledge from $T$. As illustrated in Fig.~\ref{fig:PM}, the Paraphraser Module $P$ is consist of two networks: encoder and decoder network which are stacked by several convolutional layers and transposed convolutional layers. Given the input $F_{t}^{l_{t}^{i}}$, the encoder network is used to compress $F_{t}^{l_{t}^{i}}$, while the decoder network aims to approximate $F_{t}^{l_{t}^{i}}$ as much as possible, so that the output of encoder network $P_{pm}(\cdot)$ can maintain the necessary knowledge. The encoder network $P_{pm}(\cdot)$ can be written as:
\begin{align}
{F_{t}^{l_{t}^{i}}}'=P_{pm}(F_{t}^{l_{t}^{i}})    
\end{align}
where ${F_{t}^{l_{t}^{i}}}'$ is the output of $P_{pm}(\cdot)$. In the training mode, the reconstruction loss function $\mathcal{L}_{rec}$ is:
\begin{align}
    \mathcal{L}_{rec}=\left \|F_{t}^{l_{t}^{i}}-P(F_{t}^{l_{t}^{i}}) \right \|^{2} 
\end{align}
which is utilized to minimize the discrepancy between $F_{t}^{l_{t}^{i}}$ and $P(F_{t}^{l_{t}^{i}})$. For efficient minimizing the $\mathcal{L}_{coco}$, we utilize the paraphrasing mode to avoid the phenomenon aforementioned. Consequently,the $\mathcal{L}_{coco}$ can be modified as:
\begin{align}
    \mathcal{L} _{coco}=\frac{1}{n}\sum_{i=1}^{n}
    \left \| \phi 
({F_{t}^{l_{t}^{i}}}',{F_{t}^{l_{t}^{n+i}}}') -\phi 
(F_{s}^{l_{s}^{i}},F_{s}^{l_{s}^{n+i}})\right \|_{2}^{2}
\end{align}
\subsection{Adversarial Distillation}
In addition, we introduce the adversarial distillation to align the outputs between student network and teacher network. The student network $S$ is deemed as a generator that aims to maximize the similarity between the prediction of student network $S$ and output from teacher network $T$. As shown in Fig.~\ref{fig:framework}, the discriminator $D$ is trained to distinguish whether an input, which consists of the prediction $Y_{s}/Y_{t}$ and the pathology image $X$, is from the teacher network or the student network. In the training procedure, the discriminator is supervised by the adversarial distillation function $\mathcal{L}_{ad}$,  which can be formulated as:
\begin{align}
    \mathcal{L}_{ad}=E_{Y_{t}\sim \mathcal{P}_{t}(Y_{t})}(D([Y_{t},X]))-E_{Y_{s}\sim \mathcal{P}_{s}(Y_{s})}(D([Y_{s},X])
\end{align}
where $[ , ]$ represents concatenation operation, $E(\cdot)$ indicates the expectation operation. $\mathcal{L}_{ad}$ is to minimize the wasserstein distance between $Y_{s}$ and $Y_{t}$\cite{wang2020intra}\cite{liu2019structured}. For training the generator $S$, analogously with the training strategy of cGAN, the generator is trained by $\mathcal{L}_{g}$ which is written as:
\begin{align}
    \mathcal{L}_{g}=E_{Y_{s}\sim \mathcal{P}_{s}(Y_{s})}(D([Y_{s},X]))
\end{align}
\subsection{Training Strategy}
For optimizing our proposed overall CoCo DistillNet, the training process can be classified into three steps. 
To begin with, we leverage the $\mathcal{L}_{re}$ to train $P$, the
parameters of $T$ are frozen at the same time. Next, the $\mathcal{L}_{ad}$ is employed to train the discriminator network $D$ which discriminates the input $Y_s/Y_t$ as correct as possible. At last, the conventional cross-entropy loss $\mathcal{L}_{ce}$ is combined with  $\mathcal{L}_{coco}$ and $\mathcal{L}_{g}$ to construct the fully objective $\mathcal{L}_{total}$:
\begin{align}
    \mathcal{L}_{total}=\mathcal{L}_{ce}+\lambda \mathcal{L}_{coco}-\beta \mathcal{L}_{g}
\end{align}
where $\lambda$ and $\beta$ are hyperparameters and we set $\lambda$ and $\beta$ to 1.0 and 0.1 respectively in this paper. Note that the optimizations of $D$ and $S$ are alternative implemented in the training procedure.
\begin{figure}[ht]
\begin{center}
\includegraphics[width=\linewidth]{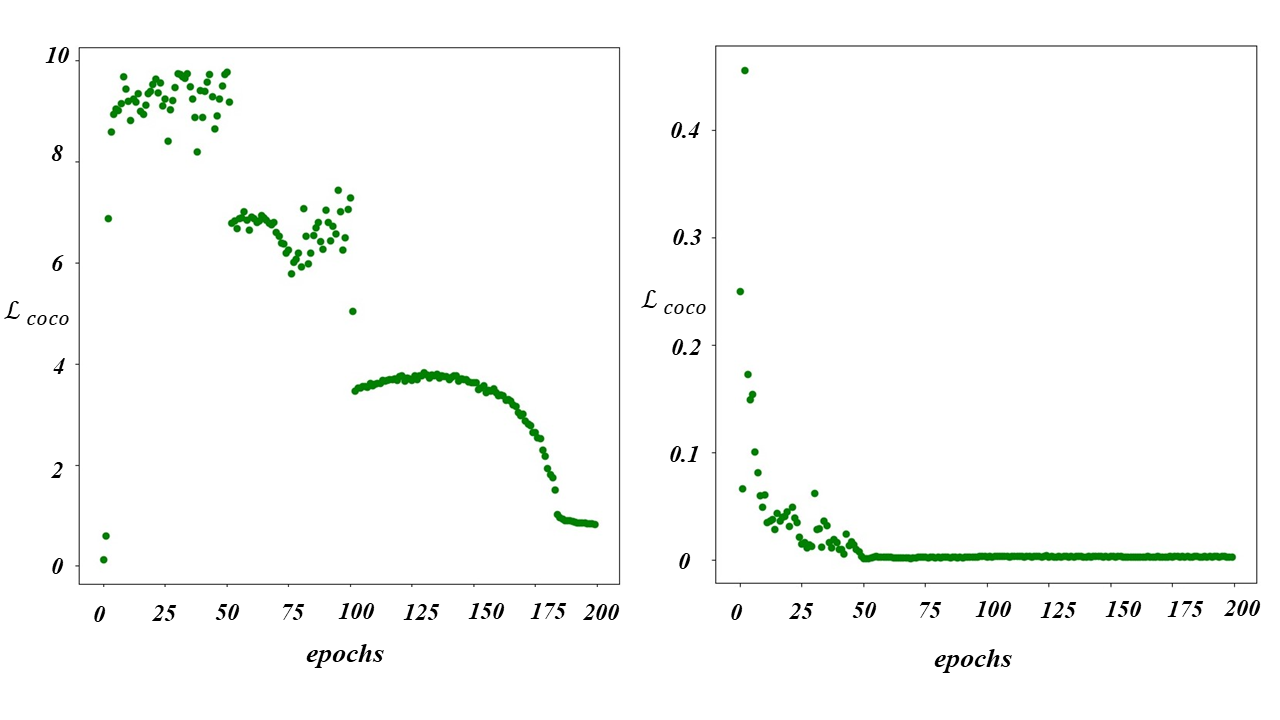}
\end{center}
   \caption{The loss curves of $\mathcal{L}_{coco}$. Left: The loss curve of $\mathcal{L}_{coco}$ without the Paraphraser Module. Right: The loss curve of $\mathcal{L}_{coco}$ with the Paraphraser Module.}
\label{fig:loss compare}
\end{figure}
\section{experiments}
\subsection{Dataset}
We employ the Gastric Cancer Segmentation Dataset\cite{sun2019accurate} to validate the effectiveness of our proposed CoCo DistillNet. The Gastric Cancer Segmentation Dataset is sampled from clinical data. There are two semantic classes in this dataset including: normal area and cancerous area. The dataset is composed of 500 pathological images with the resolution of $2048\times2048$, which are cropped from whole pathological slides of gastric regions\cite{sun2019accurate}. We adjust the training and testing images and refer to the training strategy of \cite{sun2019accurate}. In our experiments, we take patch-based method expressed as each $2048\times2048$ image is cropped into four $1024\times1024$ patches. Moreover, ordinary data augmentation methods are also employed in our training, such as randomly flip images horizontally and vertically.
\subsection{Implementation Details}
The FANet\cite{li2020accurate} which is a U-net based on self-attention (FANet) mechanism is employed as our teacher network $T$. Meanwhile, we adopt the U-net based on MobileNetV2\cite{sandler2018mobilenetv2} (Mobile U-Net) as our student network $S$. Simultaneously, the lightweight networks ENet\cite{paszke2016enet} and ERFNet\cite{romera2017erfnet} are chosen as our student network $S$ to verify the generalization of our method.

For training $S$, $T$, and $D$, we use Adam as optimizer with weight decay 0.0002 and the 'ploy' learning rate policy $(1-\frac{iter}{maxiter})^{power}$, where the $power$ is set to 0.9. For training  $P$, we employ standard SGD as optimizer with the momentum 0.9, weight decay 0.0002 and the 'poly' learning rate policy. In the process of training $S$, $T$, and $P$, the learning rate is set to 0.03 initially and divided by 10 after every 50 epochs. The initial learning rate is set to 0.0002 for optimizing $D$. All of the networks is trained for 200 epochs on the Gastric Cancer Segmentation Dataset with the batch size of 8. We resize the input images size as $512\times512$ for the limited GPU memory. All experiments are implemented on the Pytorch platform with 2 NVIDIA 2080TI GPUs. 
\subsection{Evaluation Metrics}
In all experiments, we utilize the commonly used pixel accuracy (ACC) and mean Intersection-over-Union (mIoU) to evaluate the performance of our networks. The ACC is calculated as the ratio of the correctly predicted pixels and total pixels. The mIoU is computed as the mean of the ratio of interval and union between the ground truth and the prediction of all classes. The ACC and mIoU can be formulated as:
\begin{align}
    ACC =\frac{\sum_{i=0}^{k}p_{ii}}{\sum_{i=0}^{k}\sum_{j=0}^{k}p_{ij}}
\end{align}
\begin{align}
    mIoU=\frac{1}{k+1} \sum_{i=0}^{k}\frac{p_{ii}}{\sum_{j=0}^{k}p_{ij}+\sum_{j=0}^{k}p_{ji}-p_{ii}} 
\end{align}
where $k+1$ refers to the number of classes, $p_{ii}$ represents the number of pixels which are belong to i-th class and predicted for i-th class, and $p_{ij}$ represents the the number of pixels which are belong to i-th class and predicted for j-th class. We also utilize the number of parameters (Params) and the sum of point operations (FLOPs) to measure model complexity.
\begin{table}[ht]
\begin{center}
\caption{\label{tab:ablation} Ablation study in the Gastric Cancer Segmentation Dataset SS: Spatial Similarity. CMSS: Channel Mixed Spatial Similarity module. CoCoD: CoCo Distillation. AdvD: Adversarial Distillation. CS and SS refers to only transfer the spatial similarity with or without the channel-wise information from $T$ to $S$.}
\begin{tabular}{|c|c|c|c|cc|}
\hline
\multicolumn{4}{|c|}{Method}                              & \multicolumn{1}{c|}{ACC}    & mIoU   \\ \hline
\multicolumn{4}{|c|}{T: FANet}                            & \multicolumn{1}{c|}{0.9029} & 0.8230 \\ \hline
\multicolumn{4}{|c|}{S: Mobile U-net}                     & \multicolumn{1}{c|}{0.8555} & 0.7475 \\ \hline
SS           & CMSS           & CoCoD         & AdvD          &                             &        \\ \hline
$\checkmark$ &              &              &              & \multicolumn{1}{c|}{0.8615} & 0.7567 \\
             & $\checkmark$ &              &              & \multicolumn{1}{c|}{0.8656} & 0.7630 \\
             &              & $\checkmark$ &              & \multicolumn{1}{c|}{0.8739} & 0.7761 \\
             &              & $\checkmark$ & $\checkmark$ & \multicolumn{1}{c|}{\textbf{0.8780}} & \textbf{0.7824} \\ \hline
\end{tabular}
\end{center}
\end{table}
\subsection{Ablation Study}
\emph{1) Module Ablation:}We conduct extensive experiments to verify the effectiveness of each component of our proposed CoCo DistillNet. We conduct four set of experiments for module ablation: Spatial Similarity (SS), Channel Mixed Spatial Similartiy (CMSS), CoCo Distillation (CoCoD) and Adversarial Distillation (AdvD). SS and CMSS are utilized to verify the effectiveness of our proposed Channel Mixed Spatial Similarity module. The distinction between SS and CMSS is that the channel-wise information is mixed or not mixed into the spatial similarity. CoCoD is employed to exam our proposed cross-layer correlation of channel-mixed spatial similarity. Furthermore, we apply CoCoD and AdvD to demonstrate the effectiveness of our complete method. These results are shown in Table.~\ref{tab:ablation}. From Table.~\ref{tab:ablation}, we observe that the spatial similarity improve the $S$ without distillation by $0.60\%$ (ACC) / $0.92\%$ (mIoU) and the Channel Mixed Spatial Similarity module can further increase the performance of the $S$ without distillation to $1.01\%$ (ACC) / $1.55\%$ (mIoU). It demonstrates that our proposed Channel Mixed Spatial Similarity module can  extract spatial similarity and channel information effectively. With the use of our CoCo Distillation, the improvements of the $S$ are $1.84\%$ (ACC) / $2.86\%$ (mIoU), indicating the knowledge defined as the cross-layer correlation of channel-mixed spatial similarity can significantly prompt the performance of the $S$. At last, the total CoCo DistillNet, i.e. with the use of CoCo Distillation and Adversarial Distillation, can prompt the ACC and mIoU of the $S$ to reach $87.80\%$ and $78.24\%$ respectively. Fig.~\ref{fig:efficient} shows the qualitative results of module ablation. We also demonstrate the effectiveness of the Paraphraser Module in Fig.~\ref{fig:loss compare}. We observe that the $\mathcal{L}_{coco}$ curve appears irregular vibration and significantly reduction after almost every 50 epochs without Paraphraser Module as shown in Fig.~\ref{fig:loss compare} (left). As illustrated in Fig.~\ref{fig:loss compare} (right), Paraphraser Module could effectively enhance training stabilization and converge rapidly.
\begin{table}[ht]
\begin{center}
\caption{\label{tab:kd} Comparison with different knowledge distillation methods.}
\begin{tabular}{|c|c|c|}
\hline
Method         & ACC    & mIoU   \\ \hline
S:Mobile U-Net & 0.8555 & 0.7475 \\ \hline
S+KD\cite{hinton2015distilling}           & 0.8620 & 0.7574 \\ \hline
S+AT\cite{zagoruyko2016paying}           & 0.8679 & 0.7665 \\ \hline
S+FSP\cite{yim2017gift}          & 0.8636 & 0.7600 \\ \hline
S+IFVD\cite{hou2020inter}         & 0.8682 & 0.7671 \\ \hline
S+CoCo(ours)   & \textbf{0.8780} & \textbf{0.7824} \\ \hline
\end{tabular}
\end{center}
\end{table}

\emph{2) Comparisons with other knowledge distillation methods:} As illustrated in the Table.~\ref{tab:kd}, our proposed CoCo DistillNet achieves the state-of-the-art performance compared with several popular knowledge distillation methods. Concretely, our method outperforms KD by $1.60\%$ (ACC) / $2.50\%$ (mIoU), and AT by $1.01\%$ (ACC) / $1.59\%$ (mIoU), in which KD and AT are both designed for classification task. It indicates the superiority of our method in pathology image segmentation. Compared with FSP, our methods outperforms the performance of the FSP by $1.44\%$ (ACC) / $2.24\%$ (mIoU). This is because our method constructs more appropriate cross-layer correlation than FSP. Our method also surpasses IFVD by $0.98\%$ (ACC) / $1.53\%$ (mIoU). It demonstrates that our method utilizes the structural property of pathology images and the correlation between shallow and depth layers can  effectively instruct the training of $S$, better than other methods devised for segmentation task. 
\begin{table}[ht]
\begin{center}
\caption{\label{tab:compare} Quantitative results of different student networks. MUNet represents Mobile U-Net. CoCo refers to that the student network tranined by out CoCo DistillNet.}
\begin{tabular}{|c|c|l|l|c|c|c|c|}
\hline
Type               & \multicolumn{3}{c|}{Network}     & ACC    & mIoU   & FLOPs                   & Params                  \\ \hline
T                  & \multicolumn{3}{c|}{FANet}       & 0.9029 & 0.8230 & 171.556G                & 38.250M                 \\ \hline
\multirow{6}{*}{S} & \multicolumn{3}{c|}{ERFNet}      & 0.8695 & 0.7691 & \multirow{2}{*}{3.681G} & \multirow{2}{*}{2.063M} \\ \cline{2-6}
                   & \multicolumn{3}{c|}{ERFNet+CoCo} & \textbf{0.8779} & \textbf{0.7824} &                         &                         \\ \cline{2-8} 
                   & \multicolumn{3}{c|}{MUNet}       & 0.8555 & 0.7475 & \multirow{2}{*}{1.492G} & \multirow{2}{*}{4.640M} \\ \cline{2-6}
                   & \multicolumn{3}{c|}{MUNet+CoCo}  & \textbf{0.8780} & \textbf{0.7824} &                         &                         \\ \cline{2-8} 
                   & \multicolumn{3}{c|}{ENet}        & 0.8766 & 0.7802 & \multirow{2}{*}{0.516G} & \multirow{2}{*}{0.349M} \\ \cline{2-6}
                   & \multicolumn{3}{c|}{ENet+CoCo}   & \textbf{0.8840} & \textbf{0.7919} &                         &                         \\ \hline
\end{tabular}
\end{center}
\end{table}
\subsection{Results}
\begin{figure*}[t]
\begin{center}
\includegraphics[width=\linewidth]{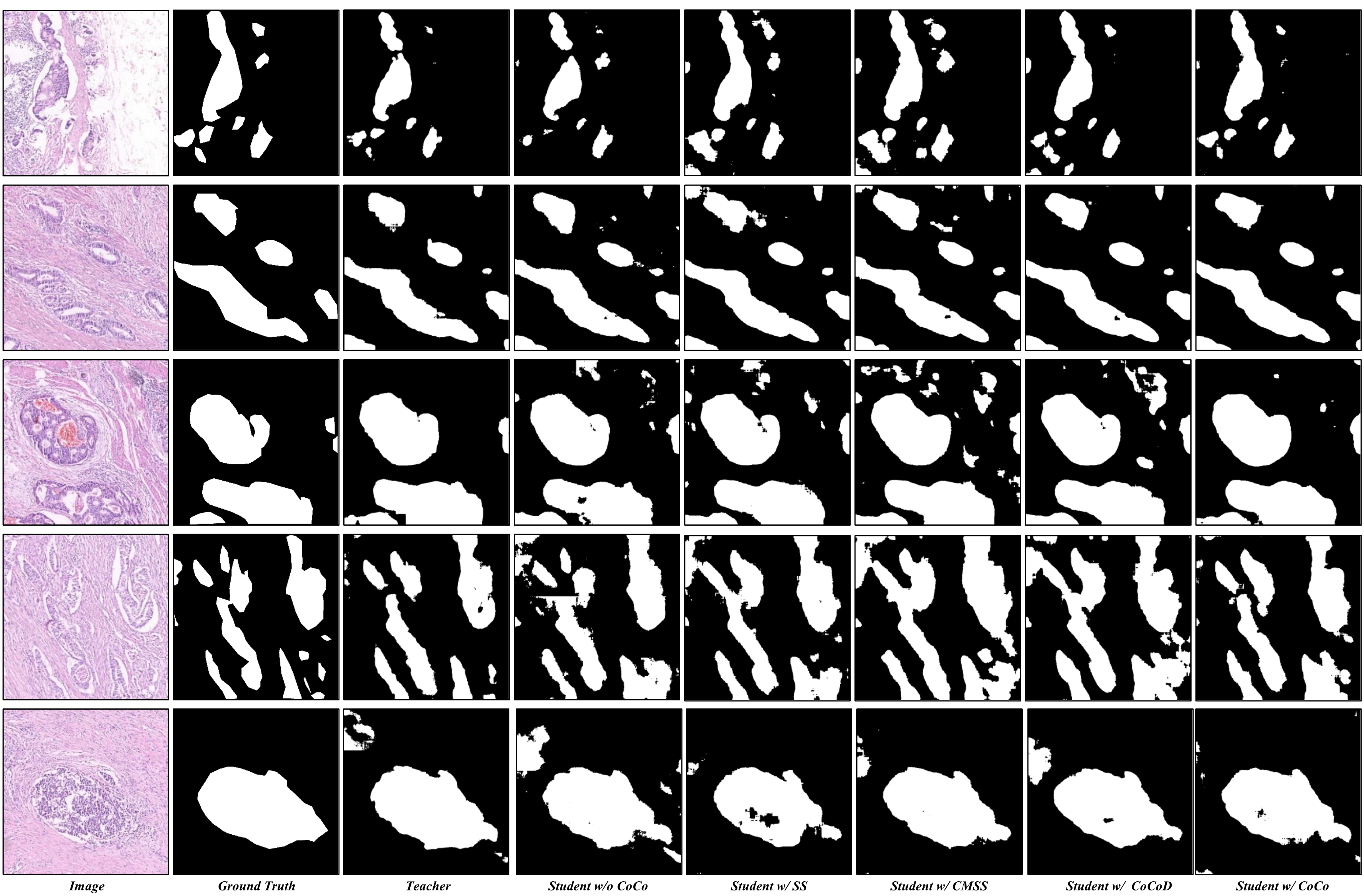}
\end{center}
   \caption{Qualitative results of our proposed method.Teacher refers to FANet. Student w/o CoCo represents the Mobile U-Net trained from scratch. SS: Spatial Similarity. CMSS: Channel Mixed Spatial Similarity. CoCoD: CoCo Distillation. CoCo: CoCo DistillNet (CoCo Distillation+Adversarial Distillation).}
\label{fig:efficient}
\end{figure*}
The effectiveness of our method is examined with several compact networks: Mobile U-Net, ERFNet, and ENet. The Table.~\ref{tab:compare} represents the performances (ACC and mIoU), complexity (FLOPs), and size (Params) of these networks. The Mobile U-Net, ERFNet, and ENet are all lightweight networks in which the Mobile U-Net has the similarity architecture with the FANet. In contrast, the ENet and ERFNet have different architecture with the FANet. As shown in Table.\ref{tab:compare}, there are  huge gaps between the performance of these lightweight networks and the cumbersome network (FANet). However, the gains of Mobile U-Net can achieve a large margin $2.25\%$ (ACC) / $3.49\%$ (mIoU) when employing our method. For ERFNet, we can see that the ACC is improved by $0.84\%$, and the mIoU is improved by $1.33\%$. At the same time, our method can boost the ENet to achieve $88.40\%$ (ACC) / $79.19\%$ (mIoU), which is a approximate performance compared with the FANet. With our proposed method, The gap between the teacher network and these compact network is alleviated significantly. It testifies that our method can improve the performance of compact networks whether the $S$ and the $T$ have similar architectures or not.

Furthermore, we also conduct qualitative analysis. As shown in Fig.~\ref{fig:efficient}, we select several pathology images with labels and show their predictions from different settings. In these experiments, the Mobile U-Net is applied as the student network $S$. We can see that the student network with our method can distinguish the cancerous regions more accurately than the student network without distillation. Moreover, our method can reduce the error of being mispredicted as cancerous regions even compared with the teacher network $T$.
\section{Conclusion}
In this paper, we present a novel knowledge distillation method tailored for pathology image segmentation. We propose the channel-mixed spatial similarity based on previous researchers. Furthermore, we define the Cross-layer Correlation of channel-mixed spatial similarity as a structural and robust knowledge. We use the knowledge transferred from the cumbersome network $T$ to improve the performance of the compact network $S$. The Adversarial Distillation is employed to further improve the performance of our method. Furthermore, we also use the Praphraser Module to stabilize our training. Our method achieves the state-of-the-art performance compared with various knowledge distillation methods. In addition, we also utilize different compact networks to verify the generalization of our method. In the future, we would like to extend our method in other multi-class and multi-model medical image segmentation tasks. We also expect to generalize the channel-mixed spatial similarity to further reduce computation cost.
\section*{Acknowledgment}
This work is completed when Wenxuan Zou is an intern under the guidance of Dr. Muyi Sun and with the help of senior Xingqun Qi. This work is supported by the National Natural Science Foundation of China under Grant 61972188.

\end{document}